\documentclass[twocolumn]{autart}    

\makeatletter
\def\@journal{}
\makeatother
\usepackage{graphicx}          
\usepackage{amsmath,dsfont}
\usepackage{booktabs}          
\usepackage{txfonts}
\usepackage{mathrsfs}        
\usepackage{natbib}          
\usepackage[utf8]{inputenc}  
\usepackage{xcolor}          
\def\replycolor{black}
\def\changecolor{black}
\newcommand{\rev}[1]{{\color{\replycolor}#1}}
\newcommand{\obs}[1]{{\color{\changecolor}#1}}

\begin{document}

\begin{frontmatter}

\title{Less Conservative Adaptive Gain-scheduling Control for Continuous-time Systems with Polytopic Uncertainties} 

\thanks[footnoteinfo]{This paper was not presented at any IFAC meeting. This study was financed in part by the Coordenação de Aperfeiçoamento de Pessoal de Nível Superior –
Brasil (CAPES) – Finance Code 001, and Conselho Nacional de Desenvolvimento Científico e Tecnológico (CNPq). Corresponding author A.C. Oliveira.}

\author[a,b]{Ariany C. Oliveira}\ead{ariany@cefetmg.br},   
\author[a,c]{Victor C. S. Campos}\ead{kozttah@ufmg.br},               
\author[a,c]{Leonardo. A. Mozelli}\ead{lamoz@ufmg.br}  
\address[a]{Department of Electrical Engineering - CEFETMG - Av. Monsenhor Luiz de Gonzaga 103, 37250-000, Nepomuceno, MG, Brazil}                                        
\address[b]{Graduate Program in Electrical Engineering - Universidade Federal de Minas Gerais - Av. Ant\^onio
Carlos 6627, 31270-901, Belo Horizonte, MG, Brazil}  
\address[c]{Department of Electronics Engineering - Universidade Federal de Minas Gerais - Av. Ant\^onio
Carlos 6627, 31270-901, Belo Horizonte, MG, Brazil}  
          
\begin{keyword}                           
Adaptive control; Linear systems; Continuous-time systems; Robust control; Linear matrix inequality (LMI); 
\end{keyword}                             

\begin{abstract}                          
\rev{The synthesis of adaptive gain-scheduling controller is discussed for continuous-time linear models characterized by polytopic uncertainties. The proposed approach computes the control law assuming the parameters as uncertain and adaptively provides an estimate for the gain-scheduling implementation. Conservativeness is reduced using our recent results on describing uncertainty: i) a structural relaxation that casts the parameters as outer terms and introduces slack variables; and ii) a precise topological representation that describes the mismatch between the uncertainty and its estimate. Numerical examples illustrate a high degree of relaxation in comparison with the state-of-the-art.}
\end{abstract}

\end{frontmatter}

\section{Introduction}

\rev{Control of} uncertain models has been a relevant research topic in control theory and applications for decades \rev{\citep{Chesi:Found:2024}}. Although many approaches have been proposed, the robust control problem is still an open challenge, \rev{due to complexity, conservatism, and the sufficiency of Lyapunov's second method \citep{Meijer:Auto:2024}.}



\rev{The design of controllers for uncertain models can be approached by adaptive control and robust control.}
The first approach generally provides less conservative results, but requires that the time-varying parameters are accessible (in real-time), through measurements \citep{ruiz2022design,nguyen2018gain} or estimations \citep{cao2024finite} to design parameter-dependent control laws. This assumption is a problem for uncertain polytopic systems in which the uncertain parameters are essentially unknown for control design.  In contrast, the second approach does not require any knowledge of the time-varying parameters \obs{\citep{chang2015robust,agulhari2012lmi,rodrigues2017parameterized,felipe2021lmi}}. However, the results are more conservative, \rev{since the compensator must guarantee a satisfactory performance over a specified range of variations}. 


\rev{Motivated by the advantages of gain-scheduling control and indirect adaptive control, recently} \cite{campos2021adaptive} presented an adaptive gain scheduling scheme for uncertain polytopic systems. The gain-scheduled control law is given by a convex sum of a fixed set of controller gains, designed via LMIs, and weighted by uncertain parameters. \rev{Also, they employ an adaptive control law, with a projection term, that guarantees that the time-varying parameters are adapted in closed-loop. Although this result is less conservative than others in the literature, it still does not fully exploit the uncertainty structure.}

Recently, \rev{\cite{wang2021control,kim2024structural,campos2025relaxation} devised clever ways to consider scheduling parameters in the LMI machinery, in the context of Takagi-Sugeno models, which are closely related to gain-scheduling. In their approaches, called structural relaxations, the scheduling parameters are treated as outfactors in the LMI formulation, leading to less conservative conditions, by potentially reducing the computational burden and introducing slack matrix variables. Concurrently, in \cite{mozelli2019computational,vieira2023geometric} the topological aspects of the time-varying nature of the scheduling parameters have been explored. They propose bounding polytopes that are efficiently tractable in the LMI framework. }

In this work, we \rev{improve} the synthesis of the adaptive gain-scheduling control for continuous-time system, affected by polytopic uncertainties. \rev{By following Lyapunov stability reasoning, new LMI conditions are obtained by using a particular structural relaxation and a less conservative approach to characterize the time-varying parameters. So, we upgrade the LMI-based conditions from \cite{campos2021adaptive}, redeveloping the associated theory with the latest scientific findings, offering less conservative results.}

\textbf{Notation}. $I$ and $\mathds{1}$ denote an identity matrix and a vector of ones, respectively, of appropriate dimensions. $\star$ represents the symmetric elements in a symmetric
matrix. $He(X)=X+X^T$. $X \otimes Y$ denotes the Kronecker product. 
$\operatorname{diag}(\cdot)$ denotes a block diagonal matrix, $\mathcal{I}_r=  \left\{i \in \mathds{N} : 1 \leq i \leq r\right\}$. $M <0$ means that M is a negative definite matrix. The following matrix notation is defined:
\begin{equation*}
    [M]_{i,j }=\begin{bmatrix}
        M_{11} & \ldots  & M_{1r} \\
        \vdots & \ddots  & \vdots\\
         M_{r1} & \ldots  & M_{rr} \\
    \end{bmatrix}, \forall i,j \in \mathcal{I}_r. \\
\end{equation*}
\section{Problem Statement}
Consider the uncertain continuous-time linear system:
\begin{equation}\label{eq:system}
    \dot{x}(t) = A(\alpha)x(t) + B(\alpha)u(t), 
\end{equation}
where $x \in \mathbb{R}^{n_x}$ is the state vector, $u \in \mathbb{R}^{n_u}$ is the control input and $\alpha = [
\alpha_1 \ldots \alpha_r]^T$
is the uncertain, possibly time-varying, parameter vector. The  uncertain matrices belong to a polytopic domain given by
\begin{equation*}
    \left[ \begin{array}{cc}
    A(\alpha) & B(\alpha)
\end{array} \right] = \sum_{i=1}^r \alpha_i \left[ \begin{array}{cc}
    A_i & B_i
\end{array} \right], \quad \alpha \in \Omega_r,
\end{equation*}
\begin{equation}\label{eq:simplex_omega_r}
    \Omega_r := \left\{ \omega \in \mathbb{R}^r : \sum_{i=1}^r \omega_i = 1, \, \omega_i \geq 0, \, i\in \mathcal{I}_r \right\},
\end{equation}
where $r$ is the number of vertices of the polytope, $A_i,B_i$, for $i \in \mathcal{I}_r$ are the vertices  and $\Omega_r$ is the unit simplex.


Since $\alpha$ is unknown, several works implemented a linear parameter independent control law $u(t)=Kx(t)$, where $K_i=K$, for $i \in \mathcal{I}_r$ for uncertain systems as (\ref{eq:system}) \obs{\citep{chang2015robust,rodrigues2017parameterized, felipe2021lmi,silva2021new}}. In general, the results obtained with this control law tend to be conservative, especially when the domain of the uncertainty is large. \rev{To circumvent}  this limitation, \cite{campos2021adaptive} proposed:
\begin{equation}\label{eq:control_law}
u(t) = \sum_{i=1}^{r} \hat{\alpha}_i(t) K_i x(t) = K (\hat{\alpha}) x(t)
\end{equation}
where the control gains $K_i \in \mathbb{R}^{n_u \times n_x}$, for $i\in \mathcal{I}_r$, and the time-varying scheduling parameter $\hat{\alpha}(t) = [\hat{\alpha}_1(t) \ldots \hat{\alpha}_r(t)]^T$ are estimates of $\alpha$. \rev{The objective is to solve the following control problem:}
\begin{prob}\label{pro:problem}
    Given a polytopic uncertain system (\ref{eq:system}), determine the control gains $K_i,~i\in \mathcal{I}_r$, and the time-varying scheduling parameter $\hat{\alpha}(t)$ such that the closed-loop system 
    \begin{equation}\label{eq:closed_loop}
        \dot{x} = \sum_{i=1}^r \sum_{j=1}^r \alpha_i \hat{\alpha}_j (A_i + B_i K_j)x.
    \end{equation}
    is asymptotically stable for $\forall \alpha \in \Omega_r$.
\end{prob}
\rev{The parameters $\hat{\alpha}_j$, $j \in \mathcal{I}_r$, used in the implementation of the control law,  must be constructed by some adaptation law. To solve this problem in the LMI framework, this adaptation law must ensure that:}
\begin{equation}\label{eq:convex_alpha_hat}
\hat{\alpha}(t) \in \Omega_r \rightarrow \sum_{j=1}^r \hat{\alpha}_j(t) = 1.    
\end{equation}
\rev{The next section jointly solves Problem~\ref{pro:problem} under the constraint \eqref{eq:convex_alpha_hat}.}
\section{LMI Design Condition}

\rev{This section solves the main problem by proposing LMI based conditions. Firstly, the structural relaxation from the TS literature is adapted to the gain-scheduling framework.} 

\begin{lem}[Adapted from \cite{campos2025relaxation}]\label{lem:relaxation}
    The following inequality
    \begin{equation}\label{eq:eq29victor}
        (\alpha \otimes I)^T \Theta (\alpha \otimes I) < 0
    \end{equation}
is equivalent to:
\begin{equation}\label{eq:eq26victor}
\Theta + \mathcal{N}\Lambda(\alpha) + \Lambda^T(\alpha) \mathcal{N}^T < 0, 
\end{equation}
where $\Theta \in \mathbb{R}^{n_xr}$ , $\mathcal{N}$ is any matrix of appropriate dimension and $\Lambda(\alpha)$ is a structural matrix such that
\begin{equation}\label{eq:eq27victor}
\Lambda(\alpha)(\alpha \otimes I) = 0. 
\end{equation}
A sufficient condition to check \eqref{eq:eq26victor} is given by:
\begin{equation}\label{eq:28victor}
\Theta + \mathcal{N}\Lambda_i + \Lambda_i^T \mathcal{N}^T < 0, \quad \forall i \in \mathcal{I}_r. 
\end{equation}
\end{lem}
\rev{Secondly, the mismatch between the model and controller parameters can be defined as:} 
\begin{equation}\label{eq:null_alpha_hat}
    \Delta \alpha_i = \hat{\alpha}_i-\alpha_i \rightarrow \sum_{i=1}^{r}\Delta\alpha_i=0
\end{equation}
\rev{Given the properties \eqref{eq:convex_alpha_hat} and \eqref{eq:null_alpha_hat}, according to \cite{mozelli2019computational,vieira2023geometric}, it follows that $\Delta \alpha$ lies in a $r-1$-dimension manifold given by:}
\begin{align}\label{eq:manifold}
    \Delta \alpha \in \Pi_q &:= \text{co}(v^1,v^2,\ldots,v^q)\\
                        & = \{v^i \in\mathbb{R}^r: -1\leq v^i_k\leq 1, \mathds{1}v^i=0\}
\end{align}
\rev{Therefore the following Lemma holds:}
\begin{lem}[Adapted from \cite{vieira2023geometric}]\label{lem:manifold}
\rev{    The columns of the following matrix $H$ are coordinates for the convex envelope of region $\Pi_q$:}
\begin{equation}\label{eq:vertexH}
    H = 
    \begin{bmatrix}
    v^1,  & v^2, & \cdots, & 
    \begin{pmatrix}
    v^\ell_1\\
    v^\ell_2 \\
    \vdots \\
    v^\ell_r
    \end{pmatrix},
    \cdots,
    v^{\Delta q}
    \end{bmatrix}
\end{equation}
\rev{The dimension of $H$ is given according to:}
    \begin{equation}
\Delta q=
\begin{cases}
\frac{r!}{\left[(r/2)!\right]^2}, & \text{if } r \text{ is even}, \\[10pt]
\frac{r!}{\left[\left( \frac{r - 1}{2} \right)!\right]^2}, & \text{if } r \text{ is odd},
\end{cases} 
\end{equation}
\rev{which follows swing factorials.}
\end{lem}
The main result is presented as follows:
\begin{thm}\label{thm:theorem}
    \rev{Consider the system described by \eqref{eq:system} and controlled by the compensator defined by \eqref{eq:control_law}. Assume that}  there exist matrices $P_i=P_i^T>0$, $P_i \in \mathbb{R}^{n_x\times n_x}$ matrices $L_{ij} \in \mathbb{R}^{n_x \times n_x}$, $N \in \mathbb{R}^{n_x \times n_x}$, $X_i \in \mathbb{R}^{n_u \times n_x}$, $\mathcal{N} \in \mathbb{R}^{2n_xr(r-1)+2n_x \times 4n_xr }$ and $\mu>0$ such that the following inequality holds:
    \begin{equation}\label{eq:infinite_condition}
         \Theta+\mathcal{N}\mathscr{B}(\alpha,\Delta \alpha) +\mathscr{B}^T(\alpha,\Delta \alpha) \mathcal{N}^T < 0, 
     \end{equation}
     where
     \begin{equation}
     \Theta  = \begin{bmatrix} \label{eq:theta_big}
                \frac{1}{2}He[\hat{Q}]_{i,j} & \star \\
                 -\frac{1}{2}[\Phi]_{i,j}               & \frac{1}{2}He[\Psi]_{i,j}
            \end{bmatrix},
    \end{equation}\begin{equation}
        \mathscr{B}(\alpha,\Delta \alpha) \begin{bmatrix}\alpha \otimes I_{2n_x} \\ \Delta \alpha \otimes I_{2n_x} \end{bmatrix}=0,
    \end{equation}
    \begin{equation}
   \hat{Q}_{ij}  = \begin{bmatrix} \label{eq:Q_big}
    He(A_iN+B_iX_j)& \star \\
    P_i - N^T + \mu \left( A_i N + B_i X_j \right) & - \mu \left( N + N^T \right)
\end{bmatrix},
    \end{equation}
    \begin{equation}
    \Psi_{ij}  =\begin{bmatrix} \label{eq:Psi_Upsilon_big}
            \text{He}(B_iX_j) + L_{ij} & \star \\
            0 & 0
            \end{bmatrix} ,~~ \Phi_{ij}=\begin{bmatrix}
            L_{ij} & \star \\
            -\mu B_iX_j & 0
            \end{bmatrix},
    \end{equation}
\rev{For every initialization of \eqref{eq:control_law} that satisfies $K_i = X_iN^{-1}, i \in \mathcal{I}_r$,}
\rev{with the time-varying scheduling parameters adapted according to:} 
\begin{align}\label{eq:adaptation_law}
    \dot{\alpha}_j(t) &= g_j(t) - \frac{1}{r} \sum_{k=1}^r g_k(t), \quad \sum_{j=1}^r \alpha_j(0) = 1,\quad \gamma > 0,\\
g_j(t) &= -\gamma \sum_{k=1}^{r} \hat{\alpha}_k x(t)^T(N^{-T} L_{kj} N^{-1}+He(N^{-T} B_k K_j)) x(t), \label{eq:g}
\end{align}
\rev{then, the closed-loop system \eqref{eq:closed_loop} asymptotically converges to the origin and all closed-loop signals are bounded.}
\end{thm}
\begin{pf}
    Consider a Lyapunov candidate function
\begin{align}
V(x, \Delta\alpha) &= x^T \mathcal{P}(\alpha) x + \frac{1}{2\gamma} \sum_{j=1}^r \Delta\alpha_j^2,\\
\mathcal{P}(\alpha) &= \sum_{i=1}^{r} \alpha_i \mathcal{P}_i, \quad \mathcal{P}_i = N^{-T} P_i N^{-1}, \quad i \in \mathcal{I}_r.\label{eq:Palpha}
\end{align}
Define $x_a=\begin{bmatrix}
        x^T &
        \dot{x}^T
    \end{bmatrix}^T $ and $\tilde{\mathcal{P}}=\begin{bmatrix}
        0 & \mathcal{P}(\alpha) \\
        \mathcal{P}(\alpha) & 0
    \end{bmatrix}$.
The time derivative of $V(x,\Delta \alpha)$ is given by
\begin{equation}\label{eq:vdot}
    \begin{split}
    \dot{V}(x, \Delta \alpha) = x_a^T
        \tilde{\mathcal{P}}x_a + \frac{1}{\gamma} \sum_{j=1}^{r} \Delta \alpha_j 
        \left( g_j - \frac{1}{r} \sum_{k=1}^{r} g_k \right),
    \end{split}
\end{equation}
The adaptation law includes the projection term $-\frac{1}{r}\sum_{k=1}^{r}g_k$
which ensures that, if $\sum_{j=1}^{r}\hat{\alpha}_j(0)=1$, then $\sum_{j=1}^{r}\hat{\alpha}_j(t)=1, \forall t $. This, in turn, implies \eqref{eq:null_alpha_hat} and
\begin{equation}\label{eq:adaptation0}
    \sum_{j=1}^{r} \Delta \alpha_j \left( \frac{1}{r} \sum_{k=1}^{r} g_k \right) = 0.
\end{equation}
Using the null-term approach from \cite[Eq. (11)]{mozelli2010}  and considering the closed-loop dynamics in (\ref{eq:closed_loop}), we have that
    \begin{equation}\label{eq:vdot_finsler}
     x_a^T 
    \tilde{\mathcal{P}}
        x_a=  x_a^T 
    \left( \tilde{\mathcal{P}} + \text{He}\left(
        \begin{bmatrix}
            N^{-T} \\
            \mu N^{-T}
        \end{bmatrix}
        \begin{bmatrix}
            A^T(\alpha) + K^T(\hat{\alpha})B^T(\alpha) \\
            -I
        \end{bmatrix}^T
    \right)
    \right)x_a.
\end{equation}
Considering \eqref{eq:adaptation0} and \eqref{eq:vdot_finsler}, it is possible rewritten \eqref{eq:vdot} as
\begin{equation}\label{eq:eq18}
    \begin{aligned}
    \dot{V}(x, \Delta \alpha) & = \sum_{i=1}^{r} \sum_{j=1}^{r} \alpha_i \hat{\alpha}_j
                x_a^T Q_{ij}x_a + \frac{1}{\gamma} \sum_{j=1}^{r} \Delta \alpha_j g_j,
                                                \end{aligned}
\end{equation}
with 
\begin{align*}
Q_{ij}&=\begin{bmatrix}
    He(N^{-T}A_i + N^{-T}B_iK_j) & \star \\
    Q_{ij}^{(21)} & -\mu \left( N^{-1} + N^{-T} \right)
\end{bmatrix},\\
Q_{ij}^{(21)} &= N^{-T}P_iN^{-1} - N^{-1} + \mu \left( N^{-T}A_i + N^{-T}B_iK_j \right).
\end{align*}
In \eqref{eq:eq18} using the relation $\hat{\alpha}_j= \alpha_j + \Delta \alpha_j$ and substituting $g_j$ for \eqref{eq:g} we have
\begin{equation}\label{eq:eq19}
    \begin{aligned}
        \dot{V} = 
        & \sum_{i=1}^{r} \sum_{j=1}^{r} \alpha_i \alpha_j 
            x_a^T
            Q_{ij}
            x_a  +2\sum_{i=1}^{r} \sum_{j=1}^{r} \alpha_i \Delta \alpha_j x^T\mu K_j^T B_i^T N^{-1}\dot{x} \\
            & - \sum_{i=1}^{r} \sum_{j=1}^{r} \Delta\alpha_i \Delta \alpha_j x^T\left[N^{-T} L_{ij} N^{-1} +He(N^{-T}B_iK_j)\right]x \\
        & - \sum_{i=1}^{r} \sum_{k=1}^{r}\alpha_i \Delta \alpha_j  x^TN^{-T}L_{ij}N^{-1} x \\
    \end{aligned}
\end{equation}
We apply the variable transformation $x = Nz$ in \eqref{eq:eq19}. Then, we defining $z_a =\begin{bmatrix}
            z^T &\dot{z}^T
            \end{bmatrix}^T$ and $X_j=K_jN$. 
By using the matrix $\Psi_{ij}$ e $\Phi_{ij}$ defined in \eqref{eq:Psi_Upsilon_big}, it is possible to rewrite \eqref{eq:eq19} as
\begin{equation}\label{eq:a3}
        \dot{V}=
        z_a^T\left(\sum_{i=1}^{r} \sum_{j=1}^{r}(\alpha_i \alpha_j 
            \hat{Q}_{ij} - \Delta \alpha_i \Delta \alpha_j \Psi_{ij}- \Delta \alpha_j \alpha_i \Phi_{ij})\right)
            z_a,
\end{equation}
where $\hat{Q}_{i,j}$ is defined in \eqref{eq:Q_big}. 
Let $\xi(\alpha) = (\alpha \otimes I_{2n_x})$, $\tilde{\xi}(\Delta \alpha)=(\Delta \alpha \otimes I_{2n_x}) $. By  employing the Kronecker product in \eqref{eq:a3}, the negativeness of $\dot{V}$ is guaranteed if $\zeta^T\Theta\zeta <0 $, with $\Theta$  defined in \eqref{eq:theta_big} and $\zeta^T=[\xi(\alpha)^T~\tilde{\xi}(\Delta \alpha)^T ]$. Eq.~\eqref{eq:infinite_condition} follows from applying the structural relaxation from Lemma~\ref{lem:relaxation}, which concludes the proof.
\end{pf}
\begin{rem}
    Unlike the approach presented in  \cite{campos2021adaptive}, we avoid using several upper bounds that introduce conservativeness. Additionally, the structural relaxation includes extra decision variables, making our approach less conservative.
\end{rem}
\begin{cor}\label{cor:main}
    A sufficient condition to guarantee \eqref{eq:infinite_condition} is to verify the following LMI conditions $~\forall \ell \in \mathcal{I}_q$,$~ m \in \mathcal{I}_{\Delta q}$:
    \begin{equation}\label{eq:main_LMI}
    \Theta+ \mathcal{N}\operatorname{diag}{(\tilde{\mathscr{B}}_m,\hat{\mathscr{B}}_\ell)} + \operatorname{diag}{(\tilde{\mathscr{B}}_m,\hat{\mathscr{B}}_\ell)}^T\mathcal{N}^T<0.
    \end{equation}
\end{cor}
\begin{pf}
    First, we need to show that $\tilde{\mathscr{B}}(\alpha)\xi=0$. Adapting \cite[Eq.~32]{campos2025relaxation} we can guarantee this constraint and it follows from Lemma~\ref{lem:relaxation} that:
    \begin{equation}
        \Theta+\mathcal{N}\operatorname{diag}(\tilde{\mathscr{B}}_m,\hat{\mathscr{B}}\Delta \alpha) +\operatorname{diag}(\tilde{\mathscr{B}}_m,\hat{\mathscr{B}}(\Delta \alpha))^T \mathcal{N}^T < 0
    \end{equation}
    Now, it remains to find a finite dimension condition to deal with the dependence on $\Delta\alpha$. \cite{vieira2023geometric} indicate algorithms to evaluate polytopes bounding the convex hull of terms such as $\hat{\mathscr{B}}(\Delta \alpha)$ with \eqref{eq:manifold}. The less conservative region is provided by the vertices indicated in matrix $H$, from Lemma~\ref{lem:manifold}. Therefore it is sufficient to verify LMIs in \eqref{eq:main_LMI}, which concludes the proof.
\end{pf}
\section{Comparative example}
\begin{figure}[htb]
    \centering
    \includegraphics[width=3in]{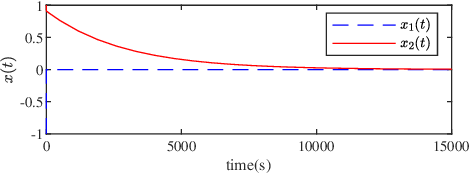}
    \caption{Simulation results under $k=775$.} \label{fig:states}
\end{figure}

\begin{figure}[htb]
    \centering
    \includegraphics[width=3in]{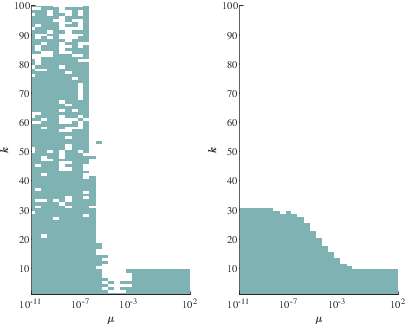}
    \caption{Feasibility regions for $\mu \in [10^{-11}, 10^{2}]$ and $k \in [1, 100]$ obtained in Example 1. Blue and white areas indicate points where LMI are feasible or infeasible, respectively. The left plot represents the conditions in Theorem \ref{thm:theorem} and the right plot represents the results in \citet[Theorem 1]{campos2021adaptive}.} \label{fig:feasibility_regions}
\end{figure}

In this section, to illustrate and verify the improvements of the proposed approach, we consider the same example presented in \cite{campos2021adaptive}. 
Consider the following uncertain polytopic system,
\begin{equation} \label{eq:system_example}
\dot{x} = \sum_{i=1}^{4} \alpha_i(A_i x + B_i u), \quad \alpha \in \Omega_4,
\end{equation}
where $k$ is a positive parameter and:
\begin{align*}
A_1 &= \begin{bmatrix} -8.1818 & 0 \\ 0.0909 & 0 \end{bmatrix}, ~ A_2 = \begin{bmatrix} -1.6364 & 0 \\ 0.0909 & 0 \end{bmatrix}, ~ A_3 = \begin{bmatrix} \frac{10(k-1)}{k+1} & 0 \\ \frac{k}{k+1} & 0 \end{bmatrix},\\
A_4 &= \begin{bmatrix} \frac{2(k-1)}{k+1} & 0 \\ \frac{k}{k+1} & 0 \end{bmatrix},~B_1 = \begin{bmatrix} -18.1818 \\ 0.0909 \end{bmatrix},~B_2 = \begin{bmatrix} -3.6364 \\ 0.0909 \end{bmatrix}, \\
B_3 &= \begin{bmatrix} -\frac{20}{k+1} \\ \frac{k}{k+1} \end{bmatrix},~ B_4 = \begin{bmatrix} -\frac{4}{k+1} \\ \frac{k}{k+1} \end{bmatrix}.
\end{align*}
The goal is to determine the largest feasible value of $k > 0$, denoted by $k^{\ast}$, for which a control solution can be found for system \eqref{eq:system_example}. \rev{Using the same tuning parameter $\mu = 10^{-11}$ for both approaches}, the largest $k$ obtained with Cor.~\ref{cor:main} is $k^*=775$ whereas with  \cite{campos2021adaptive} is $k^*=30.32$. 
\rev{Using the same basis of comparison, the improvement of the proposed approach is clear. The controller design for the system with} $k^\ast=775$ \rev{was simulated and the results are shown in Figure \ref{fig:states}}. 
\rev{Another comparison is shown in Figure \ref{fig:feasibility_regions}. In this case, Cor.~\ref{cor:main} was tested for several pairs of tuning scalar $\mu$ and parameter $k$.} Observe that Cor.~\ref{cor:main} leads to less conservative results in this case because the corresponding feasibility region is larger than that obtained in \cite{campos2021adaptive}. \rev{Also, for every value of k there exists one or more values of $\mu$ that provide a feasible controller.} For this example, we highlight that the method in \cite{campos2021adaptive} has already overcome the existing linear robust control approaches.  

\section{Conclusion}
This paper investigated the design of adaptive gain-scheduling state feedback controllers for uncertain polytopic systems. The main advantage of the proposed approach is avoiding the use of upper bounds used in the state-of-the-art, which are sources of conservatism. Combining the Lyapunov formalism,  structural relaxation approach, and slack matrices variables, an LMI-based condition is obtained such that the closed-loop is asymptotically stable and all closed-loop signals are bounded. The superiority of the proposed approach was illustrated through a numerical example.

\bibliographystyle{apalike}     
\bibliography{autosam}           

\end{document}